\documentclass[useAMS,usenatbib]{mn2e}

\usepackage[pdftex]{graphicx}
\usepackage{amssymb}
\usepackage{amsmath}
\usepackage{amstext}
\usepackage{multirow} 
\usepackage{lscape}

\newcommand{\adfo}{${\rm ADF(O^{2+})}$}
 
\newcommand{\te}{T$_{e}$}
\newcommand{\nel}{n$_{e}$}

\newcommand{\chb}{c(H$\beta$)}
\newcommand{\tf}{$t^{2}$}
\newcommand\ion[2]{#1~{\sc {#2}}\relax}        % ej. \ion{O}{ii} --> OII
\newcommand\ioni[2]{${\rm #1^{#2}}$}           % ej. \ioni{O}{+} --> O^+

\newcommand{\cmc}{{\rm cm$^{-3}$}}

\newcommand{\hii}{H~{\sc ii}}

\title[Integral field spectroscopy in the Orion Nebula]{Integral field spectroscopy of selected areas of the 
Bright Bar and Orion-S cloud in the Orion Nebula%
			       \thanks{Based on observations collected at the Centro Astron\'omico 
			       Hispano Alem\'an (CAHA) at Calar Alto, operated jointly by the 
			       Max-Planck Institut f\"ur Astronomie and the Instituto de 
			       Astrof{\'{\i}}sica de Andaluc{\'{\i}}a (CSIC).}}
\author[A. Mesa-Delgado et al.]
       {A. Mesa-Delgado\thanks{E-mail: amd@ifa.hawaii.edu/adalfis@gmail.com}$^{1,2,3}$, 
        M. N\'u\~nez-D{\'{i}}az$^{2,3}$, C. Esteban$^{2,3}$, L. L\'opez-Mart{\'{\i}}n$^{2,3}$ and
        \newauthor 
        J. Garc\'ia-Rojas$^{2,3}$ \\ 
 	$^1$Institute for Astronomy, 2680 Woodlawn Drive, Honolulu, HI 96822, USA\\       
	$^2$Instituto de Astrof\'\i sica de Canarias, E-38200 La Laguna, Tenerife, Spain \\
        $^3$Departamento de Astrof\'\i sica, Universidad de La Laguna, E-38205 La Laguna, 
	Tenerife, Spain \\
       }

\begin{document}

\date{Accepted 2011 June 16.  Received 2011 June 13; in original form 2011 May 12}
\pagerange{\pageref{firstpage}--\pageref{lastpage}} \pubyear{2010}

\maketitle
\label{firstpage}

%%%%%%%%%%%%%%%%%%%%%%%%%%%%%%%%%%%%%%%%%%%%%%%%%%%%
%%%%%%%%%%%%%%%%%%%%%%%%%%%%%%%%%%%%%%%%%%%%%%%%%%%%
\begin{abstract}
 We present integral field spectroscopy of two selected zones in the Orion Nebula obtained with the 
 Potsdam Multi-Aperture Spectrophotometer (PMAS), covering the optical spectral range from 3500 to 7200 
 \AA\ and with a spatial resolution of 1$\arcsec$. The observed zones are located on the prominent Bright 
 Bar and on the brightest area at the northeast of the Orion South cloud, both containing remarkable 
 ionization fronts. We obtain maps of emission line fluxes and ratios, electron density and temperatures, 
 and chemical abundances. We study the ionization structure and morphology of both fields, which ionization 
 fronts show different inclination angles with respect to the plane of the sky. We find that the maps of 
 electron density, \ioni{O}{+}/\ioni{H}{+} and O/H ratios show a rather similar structure. We interpret this as 
 produced by the strong dependence on density of the [\ion{O}{ii}] lines used to derive the \ioni{O}{+} 
 abundance, and that our nominal values of electron density --derived from the [\ion{S}{ii}] line ratio-- may be 
 slightly higher than the appropriate value for the \ioni{O}{+} zone. We measure the faint recombination lines of 
 \ion{O}{ii} in the field at the northeast of the Orion South cloud allowing us to explore the so-called abundance 
 discrepancy problem. We find a rather constant abundance discrepancy across the field and a mean value 
 similar to that determined in other areas of the Orion Nebula, indicating that the particular physical conditions 
 of this ionization front do not contribute to this discrepancy. 
\end{abstract}

%%%%%%%%%%%%%%%%%%%%%%%%%%%%%%%%%%%%%%%%%%%%%%%%%%%%
%%%%%%%%%%%%%%%%%%%%%%%%%%%%%%%%%%%%%%%%%%%%%%%%%%%%
\section{Introduction} \label{intro}
 The analysis of \hii\ regions is currently our best tool to study the physical conditions and chemical 
 abundances of the interstellar medium. Because of its proximity and high surface brightness, the Orion 
 Nebula (M42) is one of the best studied \hii\ regions of our galaxy. Although the Orion Nebula has 
 a large extension for more than one half degree, most of the radiation comes from the inner part, 
 the so-called Huygens region. This is an active star-formation region ionized by a group of four massive 
 stars known as the Trapezium cluster, where $\theta^1$Ori C \citep[O7 V;][]{simondiazetal06} is the main 
 ionizing source and, therefore, responsible of the bulk of nebular emission from the main ionization front 
 (MIF). Physical, chemical, kinematical and structural properties for this \hii\ region have been studied 
 in depth using different observational techniques \citep[$e.g.$][]{hesteretal91, poggeetal92, walteretal92, 
 estebanetal98, ballyetal00, odell01, odelletal03, doietal04, estebanetal04, garciadiazhenney07, 
 garciadiazetal08, odellhenney08,odelletal09} as well as photoionization models \citep[$e.g.$][]{zuckerman73, 
 balicketal74,rubinetal91,baldwinetal91,wenodell95,odell01}. The presence of many different morphological 
 features such as protoplanetary discs, Herbig-Haro objects or bright bars makes the Orion Nebula the 
 best laboratory for analysing photoionization phenomena. 
 %%%%%%%%%%%%%%% 
  \begin{figure*}
   \centering
   \includegraphics[scale=2.5]{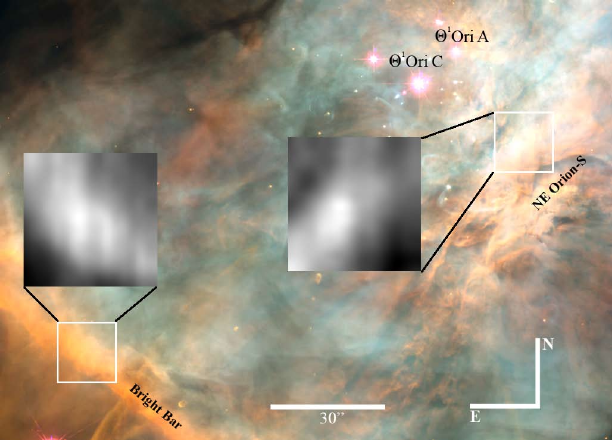} 
   \caption{$HST$ image of the central part of the Orion Nebula, which combines WFPC2 images taken 
               in different narrow-band filters \citep{odellwong96}. The white squares correspond to the fields 
               of view of PMAS IFU used, covering the Bright Bar and the northeast edge of the Orion-S 
               cloud (NE-Orion-S). The black and white images show enlargements of the H$\alpha$ emission 
               in our PMAS fields, rebinned to 160$\times$160 pixels and smoothed using a Gaussian filter.}
   \label{f1}
  \end{figure*}
 %%%%%%%%%%%%%%% 
  
 They study of ionization fronts is one of these phenomena of special interest because they are transitions 
 between the fully ionized gas and its neutral surroundings, where we expect density and temperature 
 gradients. Probably, there is an ionization front in almost every direction towards the Orion Nebula, though 
 only certain geometries allow us to explore the ionization structure of these features. The clearest example in 
 the Orion Nebula is the well-known Bright Bar (hereinafter, BB). It is one of the most conspicuous features of 
 the Orion Nebula, which is seen as an elongated structure in both ionized gas and molecular emission. It is  
 located at about 111$\arcsec$ to the southeast of $\theta^1$Ori C. The BB appears specially bright 
 in emission lines of low ionization species such as [\ion{N}{ii}] or [\ion{S}{ii}] \citep[$e.g.$][]
 {hester91,poggeetal92,walteretal93, garciadiazhenney07}. On the basis of this fact, \cite{balicketal74} 
 proposed that the BB is an escarpment in the MIF. Multi-wavelength studies from infrared to radio 
 spectral domains as well as theoretical modeling \citep[$e.g.$][and references therein]
 {tielensetal93, tauberetal94, vanderwerfetal96, marconietal98, youngowletal00, kassisetal06, 
 vanderwieletal09, pellegrinietal09} support this proposal, where the MIF changes from a 
 face-on to an edge-on geometry. \cite{pellegrinietal09} modeled the emission spectra along the transition 
 from \ioni{H}{+} to \ioni{H}{0} and H$_2$ regions. They adopt a thin slab geometry for the BB, which 
 extends 0.115 pc and is inclined 7$^\circ$ with respect to the line of sight, finding a good agreement 
 between the computed optical spectrum and that observed by \cite{baldwinetal91}.
 
 Besides the BB, there are other remarkable areas in the Orion Nebula. Recently, \cite{garciadiazhenney07} 
 have identified several examples of compact bars toward the core of the Huygens region, in addition to the 
 brighter bars previously reported in the literature \citep{odellyusefzadeh00}. This is the case of the brightest 
 zone of the Huygens region (see Fig.~\ref{f1}), located at the northeast at the northeast edge of the Orion 
 South (Orion-S) cloud (hereinafter, NE-Orion-S). This region is located about 30$\arcsec$ to the southwest 
 of $\theta^1$Ori C. In several observational studies \citep[$e.g.$][]{poggeetal92, walteretal93, 
 garciadiazhenney07} the NE-Orion-S appears as an extended structure, which is very bright in [\ion{N}{ii}] 
 and [\ion{S}{ii}] emission lines as in the case of the BB. Recently, \cite{odelletal09} have studied in-depth the 
 three dimensional structure of the whole Orion-S concluding that it is a cloud ionized from one side facing to $
 \theta^1$Ori C and suspended within the main body of the Orion Nebula in front of the MIF. Therefore, 
 this cloud as well as its NE-Orion-S edge are ionization fronts with a given unknown inclination, though it 
 should not be as tilted as the BB \citep{odelletal09}. 
 %%%%%%%%%%%%%
  \begin{table*}
   \centering
   \begin{minipage}{110mm}
     \caption{Journal of observations.}
     \label{jobs}
    \begin{tabular}{ccccc}
     \hline
                 &        &         &          \multicolumn{2}{c}{Exposure time (s)}\\     
         Field & $\alpha$$^a$ & $\delta$$^a$ &  $-$72$^{\circ}$ Blue & $-$68$^{\circ}$ Red\\
     \hline
                  Bright Bar  &   05$^h$ 35$^m$ 22\fs 3  &  -05$^\circ$ 24$'$ 33\farcs 0 & 4$\times$600,1$\times$10 & 1$\times$300,1$\times$10\\        
        NE-Orion-S  &   05$^h$ 35$^m$ 14\fs 5  &  -05$^\circ$ 23$'$ 36\farcs 0 & 1$\times$400,1$\times$10 & 1$\times$300,1$\times$10\\        

     \hline
    \end{tabular}
    \begin{description}
      \item[$^a$] Coordinates of the field centre (J2000.0).  
    \end{description}
   \end{minipage}
  \end{table*}
 %%%%%%%%%%%%%
 
 As we will see in the next paragraphs, there have been a number of works devoted to study the spatial 
 distributions of physical conditions and chemical abundances in the Orion Nebula, but this is the first one 
 specifically dedicated to these ionization fronts --BB and NE-Orion-S-- making use of integral field 
 spectroscopy in the optical spectral range at spatial scales of about 1$\arcsec$.  \cite{walteretal92} 
 determined radial gradients of the physical conditions from the analysis of 22 different areas of the nebula 
 obtained from long-slit spectrophotometry. \cite{poggeetal92} and \cite{walteretal93} studied the ionization 
 structure of the entire Huygens region by means of Fabry-Perot and CCD imaging, respectively. 
 \cite{poggeetal92} obtained the first electron density map, finding the maximum values precisely at the 
 NE-Orion-S position. More recently, \cite{odelletal03} carried out a high-spatial resolution map of the 
 electron temperature (derived from the [\ion{O}{iii}] line ratio) centered at the southwest of the Trapezium 
 cluster from flux-calibrated narrow-band images taken with the Hubble Space Telescope ($HST$). 
 \cite{rubinetal03} also analysed long-slit data taken with $STIS$ at the $HST$, obtaining the spatial 
 distributions of electron temperatures (derived from the [\ion{O}{iii}] and [\ion{N}{ii}] line ratios) along 4 slit 
 positions. Using long-slit spectroscopy at spatial scales of 1\farcs2 and covering the Huygens region with 5 
 slit positions, \cite{mesadelgadoetal08} also obtained the spatial distributions of emission lines fluxes, 
 physical conditions and chemical abundances, specially of those derived from faint recombination lines of 
 heavy-element ions. There are just a few works using integral field spectroscopy in the Orion Nebula. 
 \cite{sanchezetal07} obtained a mosaic of the Orion Nebula from integral field spectroscopy with a spatial 
 resolution of 2\farcs7 but with very short exposure times. The electron temperature map obtained by these 
 authors from the [\ion{N}{ii}] line ratio shows spatial variations and the electron density map is very rich in 
 substructures. There are only three additional studies available in the literature that used this observational 
 technique in small fields of the Orion Nebula, though they are focused in the analysis of certain morphological 
 structures. These are the cases of \cite{vasconcelosetal05} and \cite{tsamisetal11}, who study the 
 protoplanetary disc LV2, and \cite{mesadelgadoetal09a}, focused in the study of the prominent Herbig-Haro 
 object HH~202.
  
 In \S\ref{obsred} we describe the observations made with PMAS and the reduction procedure. In 
 \S\ref{medpmas} we describe the emission line measurements and the reddening correction of the spectra. 
 In \S\ref{phyab} we explain the calculations of physical conditions and chemical abundances. In 
 \S\ref{results} we present the spatial distributions of the flux of several emissions lines and 
 emission line ratios, physical conditions and chemical abundances. In \S\ref{discu} we discuss the 
 structure of the two ionization fronts studied as well as the so-called abundance discrepancy problem. 
 Finally, in \S\ref{conclu} we summarize our main conclusions. 
 
%%%%%%%%%%%%%%%%%%%%%%%%%%%%%%%%%%%%%%%%%%%%%%%%%%%%
%%%%%%%%%%%%%%%%%%%%%%%%%%%%%%%%%%%%%%%%%%%%%%%%%%%%
\section{Observations and Data Reduction} \label{obsred}  
 In Fig.~\ref{f1} we show the two fields on the Orion Nebula observed in service time on 2008 December 
 19-20 at Calar Alto Observatory (Almer\'ia, Spain). Integral field spectroscopy was performed in these 
 fields using  the Potsdam Multi-Aperture Spectrophotometer \citep[PMAS,][]{rothetal2005} at the 3.5-m 
 Telescope and its standard lens array integral field unit (IFU) of 16$\arcsec\times$16$\arcsec$ field of 
 view (FoV) with a spaxel size of 1$\arcsec$$\times$1$\arcsec$. The optical range from 3500 to 7200 
 \AA\ was covered using the V600 grating and two rotator angles ($-72^{\circ}$ for the blue spectra and 
 $-68^{\circ}$ for the red one). This grating allowed us to achieve an effective spectral resolution of about 
 3.6 \AA. The total integration times for the blue and red spectra as well as the central coordinates of 
 each field are presented in Table~\ref{jobs}. Additional short exposures of 10 seconds were also taken 
 in order to avoid saturation of the brightest emission lines. Calibration images were also obtained during 
 the night: HgNe arc lamps for the wavelength calibration and continuum lamps needed to extract the 256 
 individual spectra on the CCD. The PMAS data were reduced (bias subtraction, spectra extractions and 
 flat-fielding correction) as well as wavelength and flux calibrated using the {\sc iraf}\footnote{{\sc iraf} is 
 distributed by NOAO, which is operated by AURA, under cooperative agreement with NSF.} reduction 
 package {\sc specred} \citep[see][for more details on the reduction process]{mesadelgadoetal09a}. The 
 data were calibrated using the spectrophotometric standard stars Feige 100, Feige 34, Feige 110 and 
 G~191-B2B \citep{oke90} and the error of the absolute flux calibration was about 5\%. The night 
 reports indicated the presence of thin clouds during the observations of the BB, and this could be the 
 most likely reason of a disagreement of about 15\% in the absolute flux measurements between the 
 short and long time exposures of the BB in the red range. We corrected for this effect and we do not 
 expect that the final results are substantially affected because we only use emission line ratios in our 
 analyses. This correction was also considered in the calculation of the flux measurement errors 
 (see \S\ref{lmea}) for the spectra of this range and field. In the blue range of the BB field we found a 
 difference between the short and long exposure spectra lower than 3\% (even lower than the nominal error 
 in the flux calibration). For the NE-Orion-S, the differences between the short and large exposures were 
 also lower than 3\% along the whole wavelength coverage in both blue and red spectral ranges. The 
 seeing had an average value of about 1\farcs5 during the observations. 
 
 The integral field data cubes are subject to the effects of the differential atmospheric refraction (DAR) 
 along the direction of the parallactic angle. This effect was corrected using our own {\sc idl} routines 
 \citep[similar to the previous ones written by P.T. Wallace for {\sc starlink} and][]{walshroy90} 
 implementing the method outlined by \cite{hohenkerksinclair85} \citep[see also \S3.281 of][]{seidelmann92}  
 The procedure calculates the fractional spaxel shifts for each monochromatic image with respect to a 
 given wavelength. The maximum DAR shift was obtained for the BB field reaching values of 
 $\Delta\alpha=0$\farcs62 and $\Delta\delta=1$\farcs16 between [\ion{O}{ii}] 3727 \AA\ and H$\beta$. 
 The reference wavelengths were H$\beta$ and H$\alpha$ for the blue and red data cube, respectively. 
 These emission lines were also used to align the blue and red data cubes calculating the right 
 spaxel offsets that produce the maximum correlation between both H$\beta$ and H$\alpha$ emission maps. 
 For the two fields, these offsets reached values lower than 0\farcs 3 in $\alpha$ and $\delta$. Finally, the 
 data cubes are corrected using the {\sc bilinear} function of {\sc idl}, the calculated fractional shifts and the 
 alignment offsets. As result of all this process, the maximum coincident FoV resulting in the whole 
 wavelength range is reduced to 14$\arcsec\times$14$\arcsec$. We tested the DAR correction observing the 
 alignment of localized features at different \ion{H}{i} lines finding good agreement in all cases.  
 
%%%%%%%%%%%%%%%%%%%%%%%%%%%%%%%%%%%%%%%%%%%%%%%%%%%%
%%%%%%%%%%%%%%%%%%%%%%%%%%%%%%%%%%%%%%%%%%%%%%%%%%%%
\section{Line measurements and reddening correction} \label{medpmas}
%%%%%%%%%%%%%%%%%%%%%%%%%%%%%%%%%%%%%%%%%%%%%%%%%%%%
\subsection{Line intensity measurements} \label{lmea}
\begin{figure}
   \centering
   \includegraphics[scale=0.5]{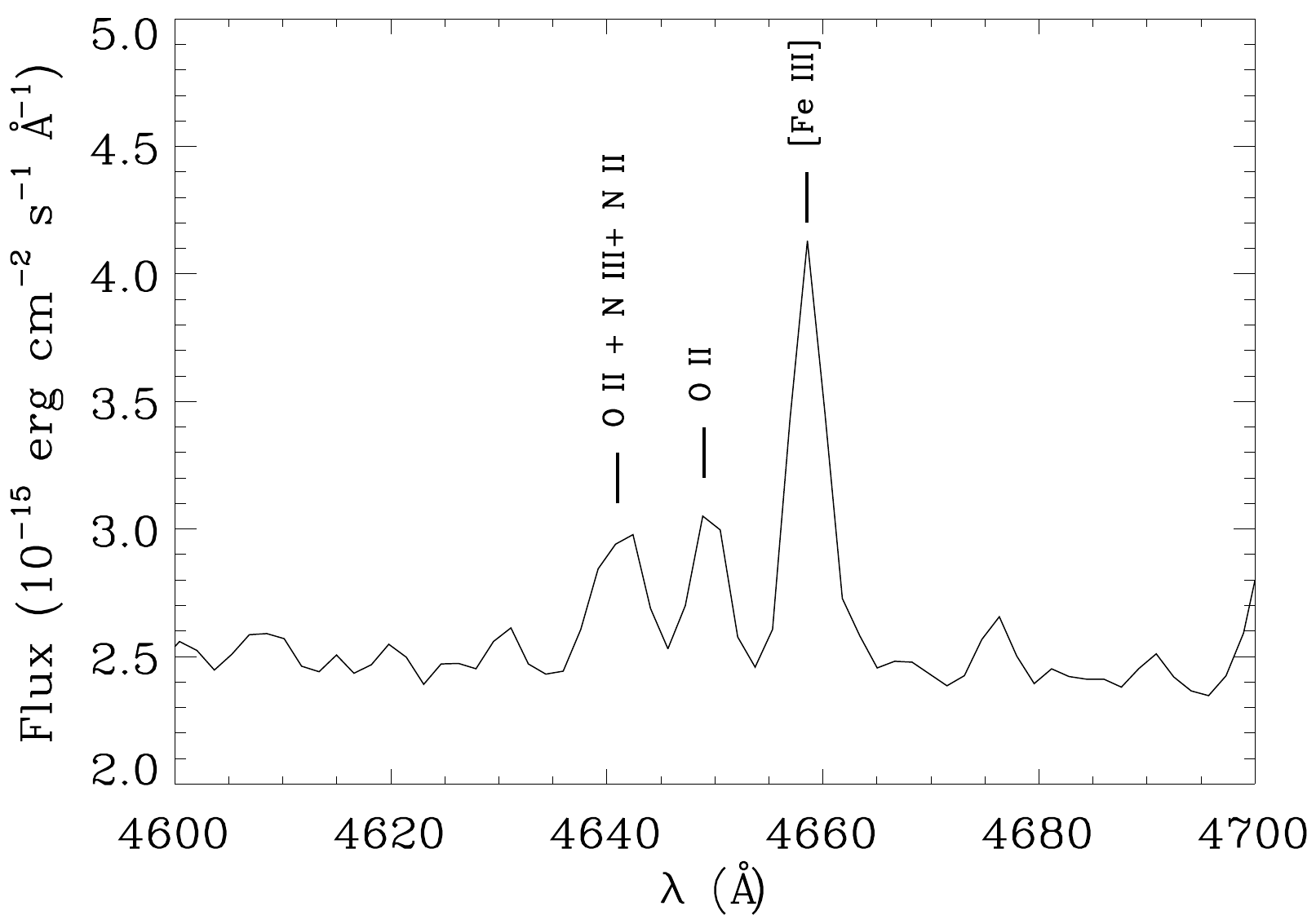} 
   \caption{Section of a PMAS spectrum around \ion{O}{ii} lines of multiplet 1 corresponding to the spaxel 
   position ($-$1:$-$2,5:6) of the NE-Orion-S field. The \ion{O}{ii} feature at about 4650 \AA\ 
   corresponds to the blend of \ion{O}{ii} 4649, 4651 \AA\ lines. The \ion{O}{ii}+\ion{N}{iii}+\ion{N}{ii} feature 
   corresponds to the blend of \ion{O}{ii} 4639, 4642 \AA, \ion{N}{iii} 4641, 4642 \AA, and \ion{N}{ii} 4643 
   \AA\ lines.}
   \label{f2}
  \end{figure}

 We measured the intensity of all the emission lines needed to perform our analysis: hydrogen Balmer lines 
 (from H$\alpha$ to H12), which are used to compute the reddening correction and verify the DAR correction; 
 collisionally excited lines (CELs) of different species (see \S\ref{dab}), which are needed to compute 
 physical conditions and chemical abundances; and, finally, the faint \ion{C}{ii} and \ion{O}{ii} lines, which 
 are used to derive the ionic abundances from recombination lines (RLs). It should be mentioned that 
 both RLs were only detected in the NE-Orion-S field. These faint lines were not detected in the BB due to its 
 lower surface brightness and lower ionization degree. In Fig.~ \ref{f2}, we present the spectra of a 
 representative spaxel, ($-$1:$-$2,5:6)\footnote{Hereinafter, the localization of an individual spaxel or 
 multiple spaxels will be given as (X1:X2,Y1:Y2) according to the coordinate system used in the maps 
 presented in next sections.}, of the NE-Orion-S, where we can see the blend of \ion{O}{ii} 4649, 4651 \AA\ 
 lines used to derive the \ioni{O}{2+}/\ioni{H}{+} ratio.  
  
 The measurement of the emission line fluxes were performed applying a single --or a multiple, in the 
 cases where it was necessary-- Gaussian profile fit procedure between two given limits and over the local 
 continuum making use of the {\sc splot} routine of {\sc iraf}. All measurements of the bright lines were 
 made with our own scripts to automatize the process. However, the flux of the faint \ion{C}{ii} 4267 \AA\ and
 \ion{O}{ii} 4650 \AA\ lines of the NE-Orion-S field were measured by hand for each individual spaxel. We have 
 determined the errors in the flux measurements following the criteria defined by \cite{mesadelgadoetal08}. 
 The final error for each emission line was the result of the quadratic sum of the flux measurement error 
 and the flux calibration error. As in our previous publications, we defined three criteria to avoid 
 spurious weak line measurements in the automatic procedure and to discriminate between real features 
 and noise: 1) line intensity peak over 2.5 times the sigma of the local continuum; 
 2) FWHM(\ion{H}{i})/1.5 $<$ FWHM($\lambda$) $<$ 1.5$\times$FWHM(\ion{H}{i}); and 
 3) F($\lambda$) $>$ 0.0001 $\times$ F(H$\beta$). 
%%%%%%%%%%%%%%%%%%%%%%%%%%%%%%%%%%%%%%%%%%%%%%%%%%%
 \subsection{Reddening correction}  \label{dchb}
  %%%%%%%%%%%%%%% 
  %%%%%%%%%%%%%%%   
  %%%%%%%%%%%%%%% 
   \begin{figure*}
   \centering
   \includegraphics[scale=0.9]{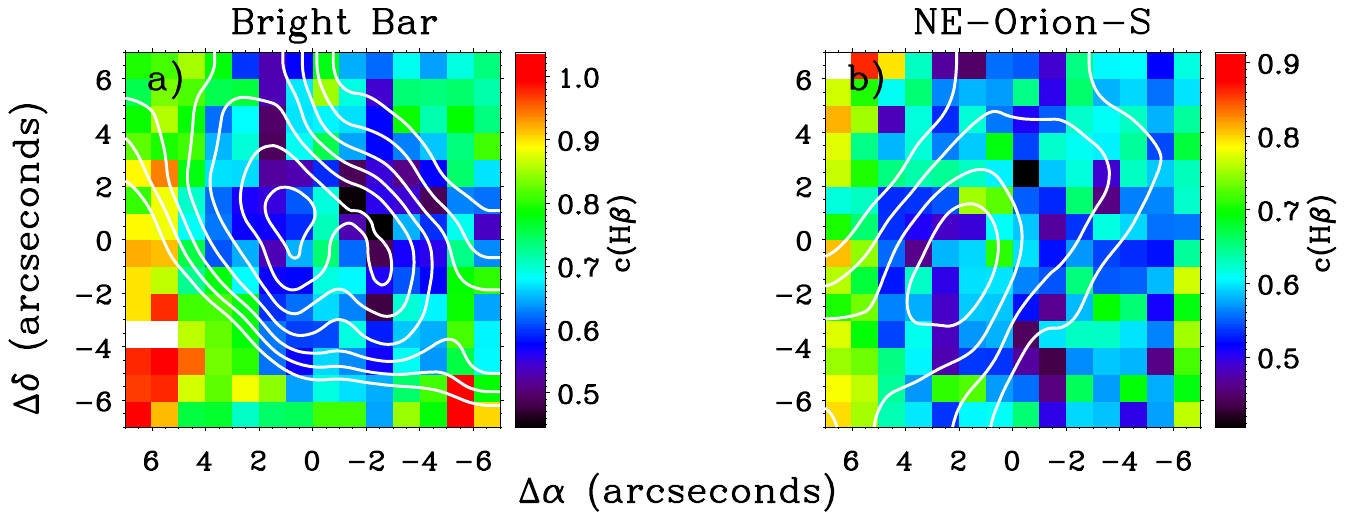} 
   \caption{Spatial distribution maps of the extinction coefficient, \chb, with H$\alpha$ contours overplotted 
                for the two PMAS fields: a) Bright Bar and b) NE-Orion-S.}
   \label{f3}
  \end{figure*}
 %%%%%%%%%%%%%%% 
 The reddening coefficient, c(H$\beta$), was obtained by fitting the observed H$\gamma$/H$\beta$ 
 and H$\delta$/H$\beta$ ratios to the theoretical ones predicted by \cite{storeyhummer95} for \nel\ = 1000 
 \cmc\ and \te\ = 10000 K. In a previous step, we considered more \ion{H}{i} line ratios in the calculations, 
 but they only added more dispersion to the final \chb\ value because the rest of available lines were either 
 rather faint (H11 and H12), or blended with other bright lines (the case of H7, H8 and H9). We used the 
 updated reddening function, $f(\lambda)$, normalized to H$\beta$ for the Orion Nebula 
 \cite[$R_V=5.5$;][]{blagraveetal07}. The use of this extinction law instead of the classical one of 
 \cite{costeropeimbert70} produces c(H$\beta$) values about 0.1 dex higher and dereddened fluxes with 
 respect to H$\beta$ about 3\% lower for lines in the range 5000 to 7500 \AA, 4\% higher for wavelengths 
 below than 5000 \AA\ and 2\% for wavelengths above 7500 \AA. The final adopted c(H$\beta$) value for 
 each spaxel is an average of the individual values derived from each Balmer line ratio weighted by their 
 corresponding uncertainties. The typical error of c(H$\beta$) is about 0.20 dex for each spaxel. All emission 
 line fluxes of a given spectrum were normalized to H$\beta$ and H$\alpha$ for the blue and red range, 
 respectively. To produce a final homogeneous set of line flux ratios, all of them were re-scaled to H$\beta$ 
 after the reddening correction. The re-scaling factor used in the red spectra was the theoretical 
 H$\alpha$/H$\beta$ ratio for the physical conditions of \te = 10000 K and \nel = 1000 \cmc.
  
 The resulting extinction maps are shown in Figs~\ref{f3}a and \ref{f3}b for the BB and NE-Orion-S, 
 respectively. The extinction coefficients vary approximately from 0.4 to 1.0 dex for the two fields. In the 
 case of the BB, the highest values, 0.9 and 1.0 dex, are located at the southeast corner of 
 the field, while the lowest values are found at the position of the ionization front in the middle of 
 the field. The NE-Orion-S presents values between 0.5 to 0.7 dex in most of the field and somewhat larger 
 values --between 0.7 and 0.9 dex-- at the east edge of the observed area. We have estimated the mean 
 \chb\ value of each field as well as the standard deviation, which amounts to 0.7$\pm$0.2 and 
 0.6$\pm$0.1 dex for the BB and NE-Orion-S, respectively. These mean values are in agreement with 
 other determinations available in the literature. \cite{odellyusefzadeh00} obtained \chb\ maps and derived 
 consistent c(H$\beta$) values both from the H$\alpha$/H$\beta$ line ratio by using calibrated $HST$ images 
 and from radio to optical surface brightness ratio. Using their extinction maps, we have found \chb\ values 
 ranging from 0.4 to 0.6 for the BB  and 0.4 to 0.7 for the NE-Orion-S at the precise positions of our fields. These 
 extinction values of \cite{odellyusefzadeh00} were determined using the classical extinction law of 
 \cite{costeropeimbert70} and, as we commented before, should be increased by about 0.1 dex to be 
 compared with our determinations. 
 
%%%%%%%%%%%%%%%%%%%%%%%%%%%%%%%%%%%%%%%%%%%%%%%%%%%%
%%%%%%%%%%%%%%%%%%%%%%%%%%%%%%%%%%%%%%%%%%%%%%%%%%%%
\section{Determinations of physical conditions and chemical abundances} \label{phyab}
%%%%%%%%%%%%%%%%%%%%%%%%%%%%%%%%%%%%%%%%%%%%%%%%%%%%
 \subsection{Physical conditions} \label{dphycon}       
  We have used the {\sc temden} task of the {\sc nebular}  package of {\sc iraf} \citep{shawdufour95} with 
  updated atomic data \citep[see][]{garciarojasetal09}, to determine the physical conditions, electron 
  densities, \nel, and temperatures, \te, for each spaxel of our PMAS fields from the usual CEL ratios 
  --[\ion{S}{ii}] 6717$/$6731 \AA\ for \nel, and [\ion{O}{iii}] (4959$+$5007)$/$4363 \AA\ and [\ion{N}{ii}] 
  (6548$+$6584)$/$5755 \AA\ for \te. For these calculations, we have followed the same procedure described 
  in \cite{mesadelgadoetal08} assuming an initial \te = 10000 K to obtain a first approximation of \nel, then 
  we calculate \te([\ion{O}{iii}]) and \te([\ion{N}{ii}]), and iterate until convergence. We have not considered 
  the contribution by recombination in the observed flux of the auroral line [\ion{N}{ii}] 5755 \AA\ to     
  derive \te([\ion{N}{ii}]), because it is very small for the conditions of the Orion Nebula 
  \citep[e.g.][]{estebanetal04}. The typical uncertainties in the density maps of the BB range from 1000 to 
  1200 \cmc. For the NE-Orion-S field, the typical error in the density distribution amounts to 4000 \cmc. This 
  higher error is due to the much higher densities at this particular area of the nebula 
  \citep[$e.g.$][]{poggeetal92,wenodell95,mesadelgadoetal08}, that makes the [\ion{S}{ii}] indicator to be 
  about the high density limit. The typical error for electron temperatures range from 500 to 1000 K for 
  \te([\ion{N}{ii}]) and from 100 to 200 K in the case of \te([\ion{O}{iii}]). 
%%%%%%%%%%%%%%%%%%%%%%%%%%%%%%%%%%%%%%%%%%%%%%%%%%%%
 \subsection{Chemical abundances} \label{dab} 
  We have derived abundances for several ions from CELs: \ioni{N}{+}, \ioni{O}{+}, \ioni{O}{2+}, \ioni{S}{+}, 
  \ioni{S}{2+}, \ioni{Ne}{2+} and \ioni{Ar}{2+} making use of the {\sc iraf} task {\sc ionic} of the package   
  {\sc nebular}. We have assumed a two-zone scheme and adopting \te([\ion{N}{ii}]) for low 
  ionization potential ions (\ioni{N}{+}, \ioni{O}{+} and \ioni{S}{+}) and \te([\ion{O}{iii}]) for high ionization 
  potential ones (\ioni{O}{2+}, \ioni{S}{2+}, \ioni{Ne}{2+} and \ioni{Ar}{2+}). We have computed the ionic 
  abundance errors as the quadratic sum of independent contributions from \nel, \te, and line flux 
  uncertainties. For ionic and total abundances of oxygen, the average uncertainties are 0.12-0.20 dex for 
  \ioni{O}{+}$/$\ioni{H}{+}, 0.03 dex for \ioni{O}{2+}$/$\ioni{H}{+} and 0.10-0.15 dex for O$/$H. 

  In addition, we have also detected and measured pure RLs of \ion{O}{ii} and \ion{C}{ii} in the majority  
  of spaxels covering the NE-Orion-S field (see Fig.~\ref{f2}). RLs have the advantage that their relative fluxes 
  with respect to \ion{H}{i} lines depend weakly on \te\ and \nel, avoiding the problem of the possible presence 
  of temperature and/or density fluctuations. These fluctuations can be actually affecting the abundance 
  determinations from CELs, whose emissivities strongly depend on the physical conditions of the ionized 
  gas \citep{peimbert67}.
  %%%%%%%%%%%%%%% 
   \begin{figure*}
   \centering
   \includegraphics[scale=0.85]{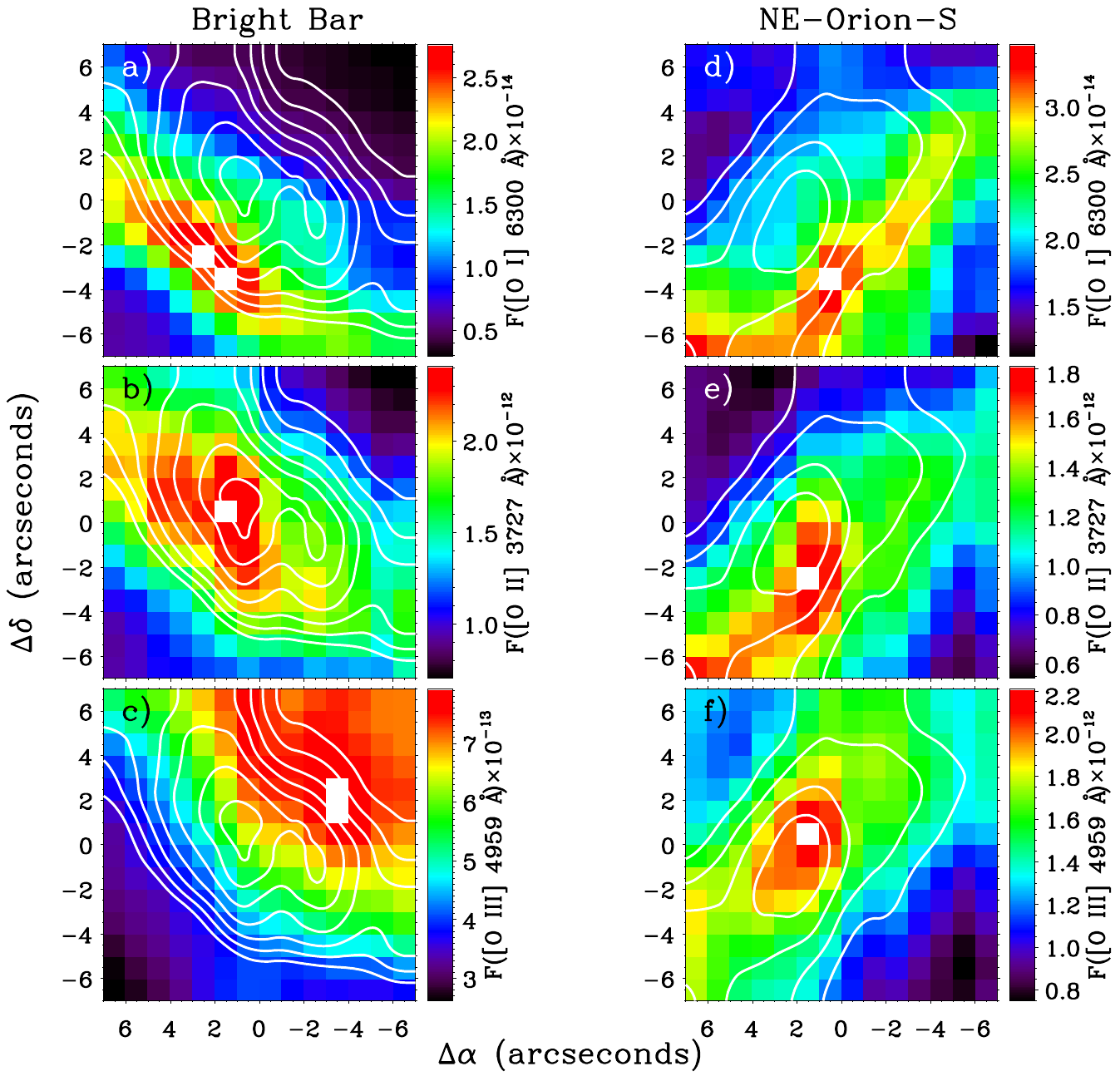} 
   \caption{Spatial distributions of emission line fluxes in units of erg~cm$^{-2}$~s$^{-1}$ of the main
                ionization stages of oxygen at the conditions of the Orion Nebula: [\ion{O}{i}] 6300 \AA\ (top), 
                [\ion{O}{ii}] 3727 \AA\ (middle) and [\ion{O}{iii}] 4959 \AA\ (bottom). The maps correspond to the 
                Bright Bar (left) and NE-Orion-S (right) fields. H$\alpha$ contours are overplotted in all 
                maps.}
   \label{f4}
  \end{figure*}
 %%%%%%%%%%%%%%% 
 
  The spectral resolution of our spectra does not allow to separate the blend of the two brightest individual 
  lines of multiplet 1 of \ion{O}{ii} at 4649 and 4651 \AA. In this situation, the combination of equations (1) 
  and (4) by \cite{apeimbertpeimbert05} should be used to determine the \ioni{O}{2+} abundance from these 
  RLs. These expressions permit to correct the line fluxes from NLTE effects in the relative intensity of 
  the individual lines of the multiplet, though they are rather small in the Orion Nebula due to its relatively 
  high density. Therefore, we have calculated the \ioni{O}{2+}$/$\ioni{H}{+} ratio from RLs as:
  \begin{equation}
    {\rm \frac{O^{2+}}{H^+}} = \frac{\lambda_{{\rm M1}}}{4861} \times %
                              \frac{\alpha_{eff}(H\beta)}{\alpha_{eff}({\rm M1})} \times %
                              \frac{I({\rm M1~O~II})}{I(H\beta)} ,
  \end{equation}    
  where $\alpha_{eff}(H\beta)$ and $\alpha_{eff}(\rm M1)$ are the effective recombination coefficients for 
  H$\beta$ and for the \ion{O}{ii} multiplet 1, respectively, and $\lambda_{\rm M1}$ = 4651.5 \AA\ the 
  representative mean wavelength of the whole multiplet. We have also calculated the 
  \ioni{C}{2+}$/$\ioni{H}{+} ratio from the flux of \ion{C}{ii} 4267 \AA\ RL and using an analogous equation 
  to (1) particularized for the case of this ion. According to the two-zone scheme, we have adopted  
  \te([\ion{O}{iii}]) to calculate the abundances of both ions, \ioni{O}{2+} and \ioni{C}{2+}. We have used 
  the effective recombination coefficients of \citealt{storey94} to derive the \ioni{O}{2+} abundances 
  (assuming LS coupling) and \citealt{daveyetal00} for \ioni{C}{2+} abundances. The typical errors in the 
  abundances from RLs amount to 0.08-0.10 dex for \ioni{C}{2+}$/$\ioni{H}{+} and 0.10-0.12 dex for 
  \ioni{O}{2+}$/$\ioni{H}{+}.  
 
%%%%%%%%%%%%%%%%%%%%%%%%%%%%%%%%%%%%%%%%%%%%%%%%%%%%
%%%%%%%%%%%%%%%%%%%%%%%%%%%%%%%%%%%%%%%%%%%%%%%%%%%%
\section{Results: spatial distribution maps} \label{results}
 In this section we analyse the spatial distributions of several emission line fluxes, emission line ratios, 
 electron densities and temperatures as well as abundances for several representative ions for each 
 PMAS field. 
 %%%%%%%%%%%%%%% 
   \begin{figure*}
   \centering
   \includegraphics[scale=0.85]{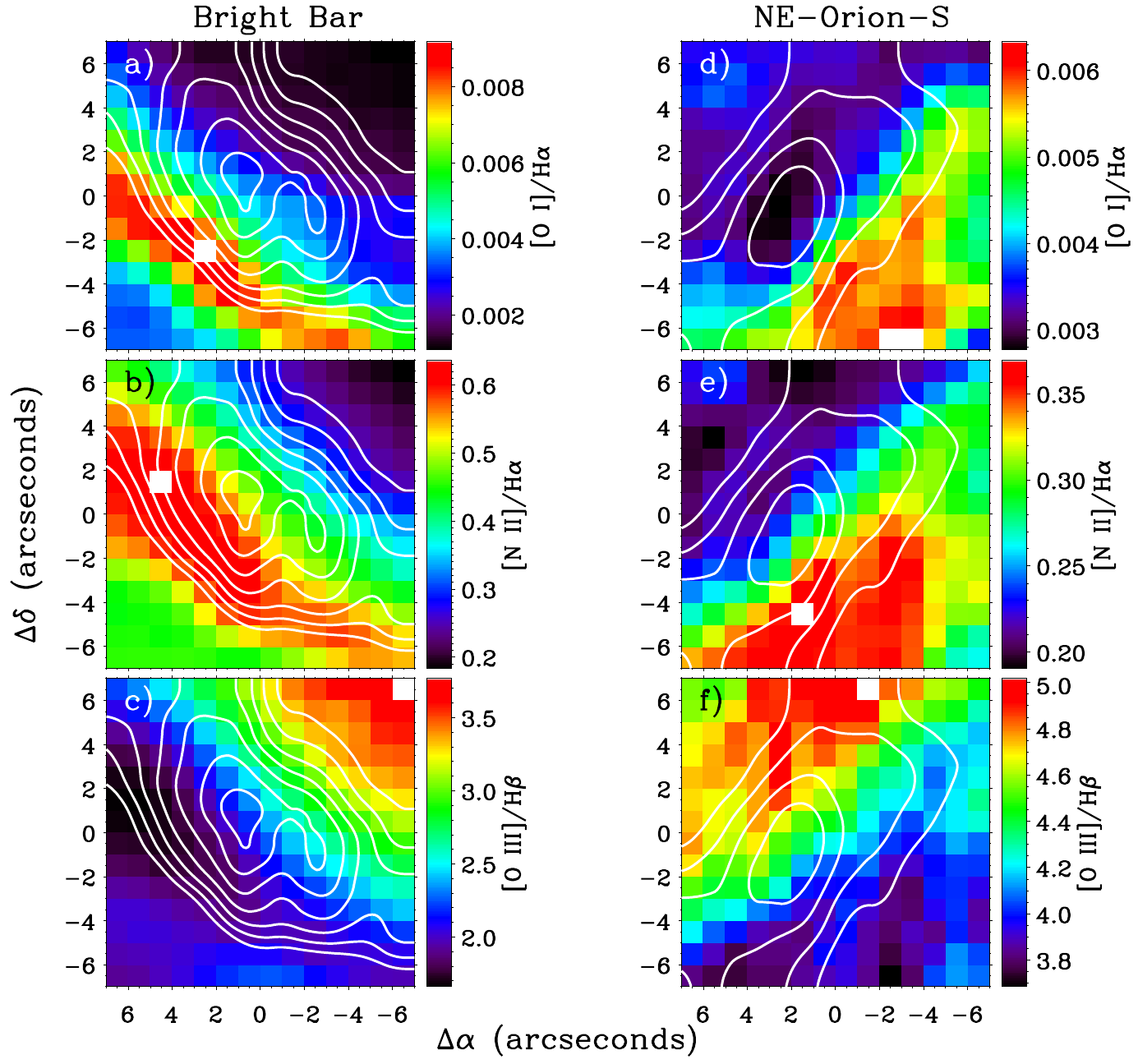} 
   \caption{Spatial distribution maps of selected emission line ratios for the Bright Bar (left) and NE-Orion-S 
                (right). The sum of [\ion{N}{ii}] 6548, 6584 \AA\ and [\ion{O}{iii}] 4959, 5007 \AA\ 
                nebular lines have been used to derive the [\ion{N}{ii}]/H$\alpha$ (middle) and [\ion{O}{iii}]/H$\beta$ 
                (bottom) ratios, respectively. H$\alpha$ contours are overplotted in all maps.}
   \label{f5}
  \end{figure*}
 %%%%%%%%%%%%%%% 
%%%%%%%%%%%%%%%%%%%%%%%%%%%%%%%%%%%%%%%%%%%%%%%%%%%%
 \subsection{Emission line fluxes and excitation ratios} \label{elf}
 In Fig. \ref{f4}, we present the spatial distribution maps of [\ion{O}{i}] 6300 \AA, [\ion{O}{ii}] 3727 \AA\ 
 and [\ion{O}{iii}] 4959 \AA\ emission line fluxes for both the BB (Figs~\ref{f4}a--\ref{f4}c) and NE-Orion-S 
 (Figs~\ref{f4}d--\ref{f4}f). These maps show the distribution of the different ionic species of oxygen that can 
 be found in the Orion Nebula: \ioni{O}{0}, \ioni{O}{+} and \ioni{O}{2+}. As it can be seen, the ionization 
 stratification of the two ionization fronts is clearly resolved in the maps. However, these maps are really 
 showing a combination of the ionization stratification and the density distribution of the gas because the 
 emissivity of a line is proportional --at first order-- to \nel$^2$. It is very illustrative to compare these 
 maps with those of the [\ion{O}{i}]/H$\alpha$, [\ion{N}{ii}]/H$\alpha$ and [\ion{O}{iii}]/H$\beta$ line ratios 
 shown in Fig.~\ref{f5}. Line ratios are directly related to the local ionization degree of the gas and do not 
 depend in a direct manner on density. In Fig.~\ref{f5} we represent [\ion{N}{ii}]/H$\alpha$ instead of 
 [\ion{O}{ii}]/H$\beta$ because the [\ion{N}{ii}]/H$\alpha$ ratio is the most commonly used indicator for the 
 analysis of ionization fronts and shows a very similar spatial distribution to [\ion{O}{ii}]/H$\beta$. As it is 
 explained in \cite{odelletal09}, the [\ion{N}{ii}]/H$\alpha$ ratio is a good indicator to detect when the MIF is 
 tilted and, therefore, this is the reason that the BB is so well defined in Fig.~\ref{f5}.

 As it has been said before, Figs~\ref{f4}--\ref{f5} show the ionization stratification at and around the 
 ionization fronts . In both figures, it is evident that the areas with the higher [\ion{O}{i}] 6300 \AA\ line 
 emission and [\ion{O}{i}]/H$\alpha$ ratio are narrower than those of the other lines represented in the figures, 
 and their maxima tend to be also located farther away with respect to $\theta^1$Ori C (see Fig.~\ref{f1}). We 
 can notice some differences between the spatial distributions of line fluxes and ratios for the BB and 
 NE-Orion-S. Firstly, the widths of the zones occupied by the tracers of the ionization fronts, [\ion{O}{i}] and 
 [\ion{N}{ii}], are narrower for the BB than for the NE-Orion-S. This indicates that the inclination angles of the 
 planes of the fronts with respect to the line of sight of both features are different. In fact, as we 
 commented in \S\ref{intro}, this was already suggested by \cite{odelletal09}. The BB is located at a border 
 of the Huygens region where the MIF is an edge-on escarpment, where its plane is tilted about 7$^\circ$ 
 with respectÊto our line of sight \citep{pellegrinietal09}. The situation is not so clear in the case of the 
 NE-Orion-S. It is a portion the Orion-S cloud, which is suspended in front of the MIF with an unknown 
 inclination and can be interpreted as a bump in the otherwise concave surface of the MIF at the 
 Huygens region \citep[][]{odelletal09}. Secondly, another difference between the BB and NE-Orion-S can 
 be seen comparing the maps shown in Figs~\ref{f4} and \ref{f5}. In the case of the BB, Fig.~\ref{f4}b) shows 
 that the maxima of [\ion{O}{ii}] 3727 \AA\ and H$\alpha$ emission coincide. However, in the case of the 
 NE-Orion-S, Fig.~\ref{f4}f) indicates that the maximum of H$\alpha$ emission coincides with [\ion{O}{iii}] 
 4959 \AA. This difference is due to the NE-Orion-S area is nearer to $\theta^1$Ori C than the BB and that 
 while few photons capable of producing \ioni{O}{2+} are reaching the BB, NE-Orion-S is illuminated by a 
 significant fraction of such high-energy photons. This is clearly confirmed in Fig.~\ref{f5}c), where the 
 maximum of the [\ion{O}{iii}]/H$\beta$ line ratio is well outside the BB and towards the ionizing source of 
 the nebula. Interestingly, Fig.~\ref{f5}f) indicates that in the case of the NE-Orion-S, the zone of the highest 
 [\ion{O}{iii}]/H$\beta$ line ratio is just at the north edge of the field. This area coincides with a filament, 
 which seems to be out of Orion-S cloud itself. Its higher excitation suggests that this filament should 
 be nearer to $\theta^1$Ori C and perhaps belongs to the MIF.
 %%%%%%%%%%%%%%% 
   \begin{figure*}
   \centering
   \includegraphics[scale=1.0]{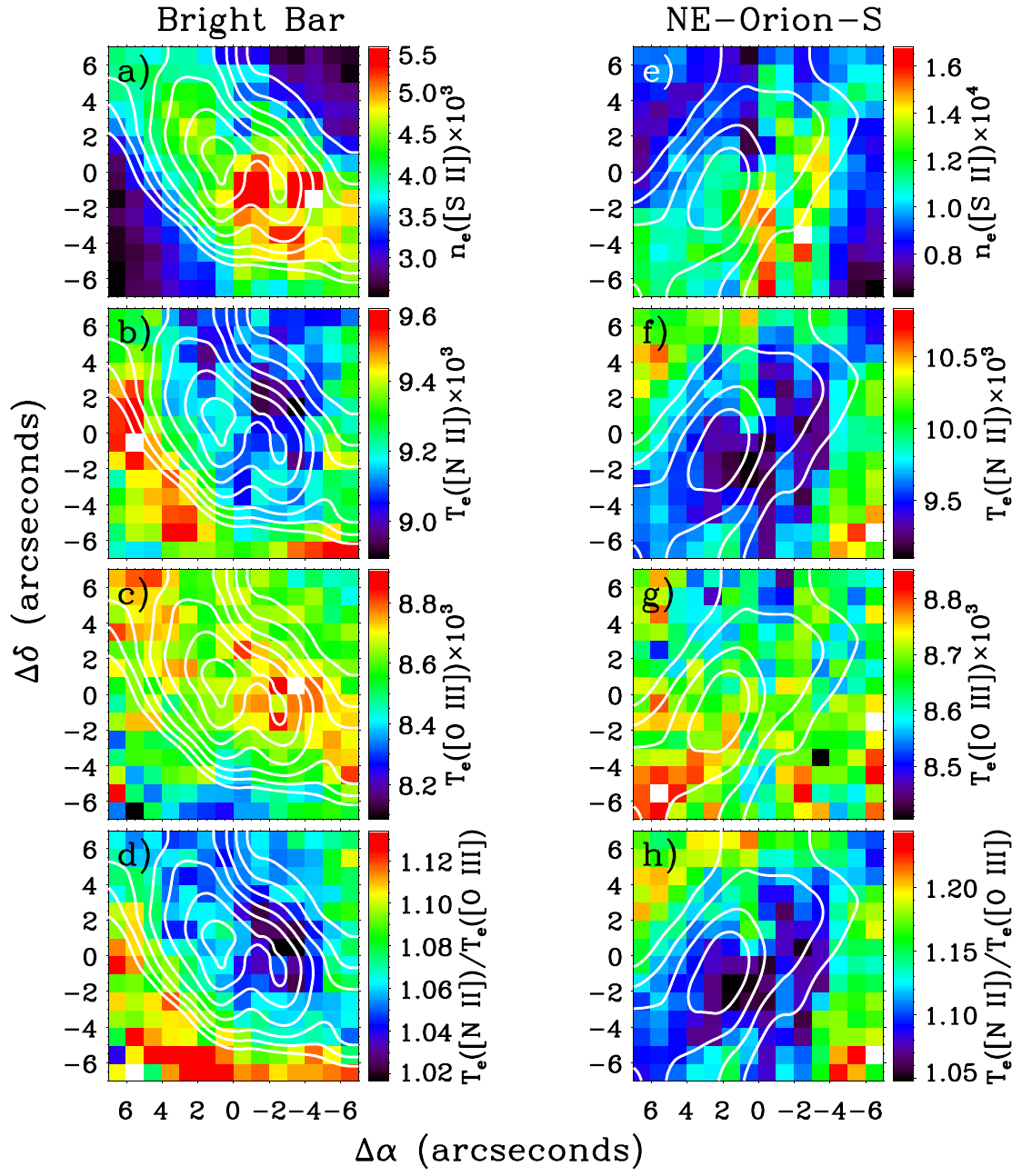} 
   \caption{Spatial distributions of physical conditions for the Bright Bar (left column) and NE-Orion-S 
                (right column). For all fields, we present \nel([\ion{S}{ii}]) (top row) in units of \cmc, 
                \te([\ion{N}{ii}]) (second row from top) and \te([\ion{O}{iii}]) (third row from top) in 
                units of K, and the ratio of both temperatures (bottom row). H$\alpha$ contours are 
                overplotted in all maps.}
   \label{f6}
  \end{figure*}
 %%%%%%%%%%%%%%% 
%%%%%%%%%%%%%%%%%%%%%%%%%%%%%%%%%%%%%%%%%%%%%%%%%%%% 
\subsection{Physical conditions} \label{phycon}
 In Fig.~\ref{f6} we present the physical conditions for the two observed fields: \nel([\ion{S}{ii}]), 
 \te([\ion{N}{ii}]), \te([\ion{O}{iii}]) and the temperature ratio \te([\ion{N}{ii}])/\te([\ion{O}{iii}]).
 
 Both fields show \nel\ spatial variations of the order of a factor 2. The densities in the BB range from 
 3000 to 5500 \cmc, but larger values are found in the NE-Orion-S, from 8000 to 16000 \cmc. In the BB field, 
 maximum densities are found within the H$\alpha$ contours, while in the NE-Orion-S these are displaced with 
 respect to the maximum of \ion{H}{i} emission. The average densities of the ionization fronts are 
 4370$\pm$580 \cmc\ for the BB and 10900$\pm$1900 \cmc\ for the NE-Orion-S. In the case of the 
 NE-Orion-S field, the highest density values are about the high density limit of the [\ion{S}{ii}] line ratio 
 indicator and, therefore, they may not correspond to the true densities. Fortunately, we have detected at 
 least 5 [\ion{Fe}{iii}] lines of the 3F multiplet ([\ion{Fe}{iii}] 4607, 4658, 4702, 4734 and 4755 \AA) and one line 
 of the 2F multiplet ([\ion{Fe}{iii}] 4881 \AA) at those spaxels of the NE-Orion-S field showing the 
 \nel([\ion{S}{ii}]) maxima. These lines permit to get an additional determination of the density, 
 \nel([\ion{Fe}{iii}]), which range of validity has an upper limit above that of the [\ion{S}{ii}] indicator. To carry 
 out these calculations we have used a 34-level atom using the collision strengths from \cite{zhang96} and the 
 transition probabilities of \cite{quinet96} and \cite{johanssonetal00}. The calculations of \nel([\ion{Fe}{iii}]) give 
 densities varying from 10000 to 18000 \cmc, even though they have large uncertainties (between 3000 and 
 6000 \cmc) due to the faintness of the [\ion{Fe}{iii}] lines. The differences among the \nel([\ion{Fe}{iii}]) and 
 \nel([\ion{S}{ii}]) determined for a given spaxel amount from 1\% to 30\%, and it is clear that 
 these differences are within the uncertainties of both indicators. Therefore, the high \nel([\ion{S}{ii}]) we 
 obtain in Fig.~\ref{f6}e) should be considered at least close to the true electron density of the ionized gas 
 at the NE-Orion-S (further discussion of these results can be found in \S\ref{cheab}).
  
 The spatial distributions of \te([\ion{N}{ii}]) show a similar qualitative behaviour in both ionization fronts 
 (Figs.~\ref{f6}b and \ref{f6}f). The pattern seems to be almost inverse to the density distributions reaching 
 the minimum temperatures approximately at the maximum densities. In the case of the BB, the range of 
 variation of \te([\ion{N}{ii}]) along the field is rather narrow reaching the maximum temperatures --about 
 9500 K-- just at the outer edge of the bar. In the case of the NE-Orion-S field, the range of values of 
 \te([\ion{N}{ii}]) is wider and the temperatures are always higher than in the BB. The maximum values 
 of temperature are also found at the outer edge of the ionization fronts. 

 The spatial distributions of \te([\ion{O}{iii}]) (Figs~\ref{f6}c and \ref{f6}g) are rather featureless with mean 
 values of about 8620 and 8670 K and variations of the order 600 and 400 K in the BB and NE-Orion-S fields, 
 respectively. In both fields, the maximum values of \te([\ion{O}{iii}]) tend to be located within the 
 \ion{H}{i} emission. In the NE-Orion-S, we can see a zone of higher values located at the southwest corner of 
 the field, though these values have also higher uncertainties --about 200-250 K. 
 
 In Figs~\ref{f6}d) and \ref{f6}h) we present the \te([\ion{N}{ii}])/\te([\ion{O}{iii}]) ratio for both fields. As we 
 can see, \te([\ion{N}{ii}]) is always slightly higher than \te([\ion{O}{iii}]). This is the expected behaviour 
 in an ionization bounded nebula as a consequence of the hardening of the incident radiation field due to
 photoelectric absorption in the low ionization zones where the [\ion{N}{ii}] lines are emitted. This ratio tends 
 to be higher in the outer zones of both fronts. 

 We have compared our determinations of the physical conditions with previous ones available in the 
 literature for the same or nearby areas. These are the cases of: the long-slit positions 1, 3, 4 and 5 
 analysed by \cite{mesadelgadoetal08}, which cover a small area of the Orion-S cloud and the BB; the 
 long-slit positions 2 and 4 studied by \cite{rubinetal03}, which cover some parts of both ionization fronts; 
 several of the slit positions observed by \cite{baldwinetal91}; the results presented by \cite{poggeetal92} for 
 the whole Huygens region; and the results from the model of \cite{wenodell95}. In general, we have found a 
 good agreement between the results obtained in these papers and our determinations, reproducing some 
 of the patterns discussed in the previous paragraphs. However, we have found a slight disagreement 
 with the \nel\ determinations of \cite{poggeetal92} for the BB. These authors analysed a spatial profile of 
 \nel\ crossing the BB in an area located about 40$\arcsec$ to the southwest of our field and with a position 
 angle of 150$^\circ$. Their profile reaches densities of about 3200 \cmc --1000 \cmc\ lower than our 
 determination-- but in agreement with the results by \cite{garciadiazhenney07} along a similar slit position. 
 On the other hand, our density determination is in agreement with the model results of \cite{wenodell95}. 
 In their figure 4a, they present a spatial profile of the density with a position angle of 322$^\circ$, 
 which precisely crosses our BB field, finding densities of about 5000 \cmc, similar to our values. Therefore, 
 the small differences found here between our physical conditions and those reported by other authors seem 
 to be related to the presence of local variations of the physical conditions in the BB. \cite{wenodell95}, 
 in their figure 5a, also presented a spatial profile of the density with a position angle of 226$^\circ$ covering 
 the NE-Orion-S field, which passes close to the southeast corner of our field. They found densities no 
 higher than 12000 \cmc, slightly lower than our determinations but consistent within the uncertainties.
 %%%%%%%%%%%%%%% 
   \begin{figure*}
   \centering
   \includegraphics[scale=0.85]{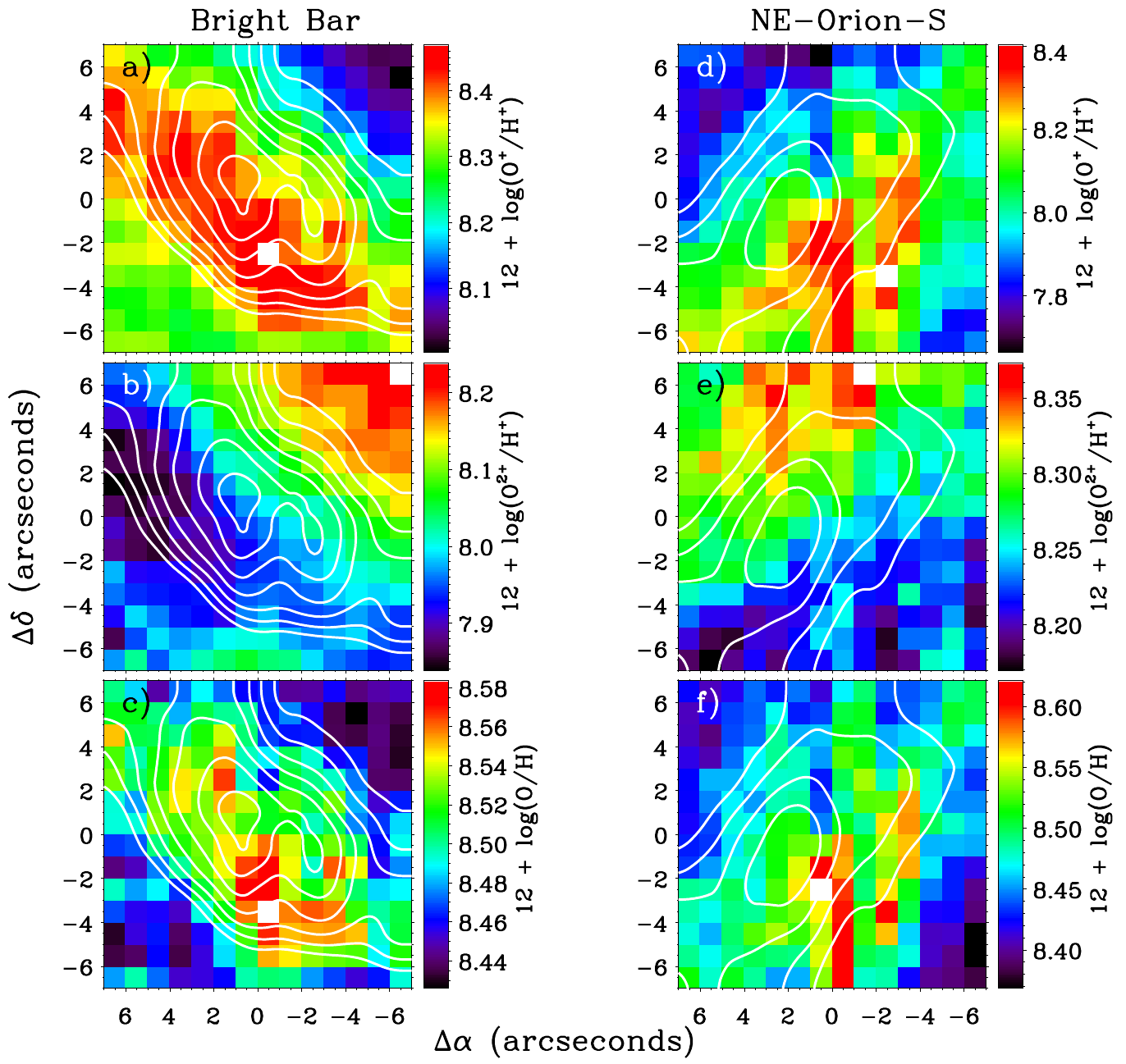} 
   \caption{Spatial distributions of the \ioni{O}{+}$/$\ioni{H}{+}, \ioni{O}{2+}$/$\ioni{H}{+} and O$/$H ratios 
   	        --in the usual logarithmic scale-- for the Bright Bar (left column) and NE-Orion-S (right 
	        column). H$\alpha$ contours are overplotted in all maps.}
   \label{f7}
  \end{figure*}
 %%%%%%%%%%%%%%% 
 %%%%%%%%%%%%%%% 
   \begin{figure*}
   \centering
   \includegraphics[scale=0.85]{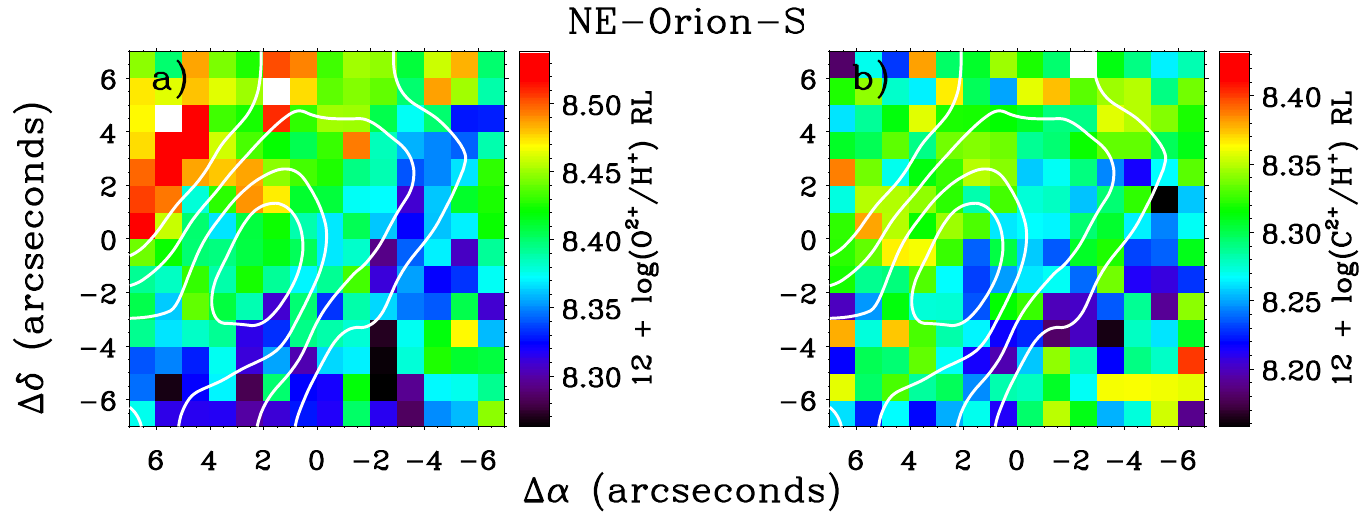} 
   \caption{Spatial distributions of \ioni{O}{2+} and \ioni{C}{2+} abundances determined from 
                recombination lines for the NE-Orion-S. H$\alpha$ contours are overplotted.}
   \label{f8}
  \end{figure*}
 %%%%%%%%%%%%%%% 
%%%%%%%%%%%%%%%%%%%%%%%%%%%%%%%%%%%%%%%%%%%%%%%%%%%%
 \subsection{Chemical abundances}  \label{cheab}
 In Fig.~\ref{f7} we present the spatial distributions of ionic and total oxygen abundances derived from 
 CELs for the two fields. As expected, the \ioni{O}{+} and \ioni{O}{2+} abundance distributions are 
 rather similar to those of the [\ion{N}{ii}]/H$\alpha$ and [\ion{O}{iii}]/H$\beta$ line ratios shown in 
 Fig.~\ref{f5}. However, unexpectedly, we find structures in the spatial distributions of the total O 
 abundance of both fields as we can see in Figs~\ref{f7}c) and \ref{f7}f). The mean values of the 
 O/H ratio at the BB and NE-Orion-S amount to 8.49$\pm$0.03 dex and 8.48$\pm$0.05 dex, 
 respectively, which are in complete agreement with the typical oxygen abundances --8.50 dex-- obtained 
 by previous works in different parts of the Orion Nebula \citep{estebanetal98, estebanetal04, 
 blagraveetal06, mesadelgadoetal09b, simondiazstasinska11}. In the BB field, the O abundance variations 
 are within the typical error of about 0.10-0.15 dex, while in the NE-Orion-S the range is slightly wider, finding 
 the lowest values at the southwest corner of the field, precisely where the \ioni{O}{2+} abundances are 
 lower and \te([\ion{O}{iii}]) has a larger uncertainty, as we mentioned in \S\ref{phycon}. In Fig.~\ref{f7}, we 
 can see that the areas with the highest total O abundance coincide in general with those showing higher 
 \ioni{O}{+}/\ioni{H}{+} ratio. Moreover, the structure of the spatial variations of the O/H ratio shown in 
 Figs~\ref{f7}c) and \ref{f7}f) is remarkably similar to that of the \nel\ maps of Figs~\ref{f6}a) and \ref{f6}e). 
 To understand this behaviour we have to consider that ionic abundance determinations derived from CELs 
 depend very much on the physical conditions of the gas. \ioni{O}{+}/\ioni{H}{+} and \ioni{O}{2+}/\ioni{H}{+} 
 ratios are both strongly dependent on \te, but not on \nel\ at the same level. In particular, the 
 \ioni{O}{+}/\ioni{H}{+} ratio derived from the [\ion{O}{ii}] 3727 \AA\ doublet is much more dependent on 
 \nel\ than the \ioni{O}{2+} abundance calculated from the nebular [\ion{O}{iii}] 4959, 5007 \AA\ lines
 at least at the densities measured in our PMAS fields. Therefore, the striking similar structure of the O/H 
 and \nel\ maps may be produced by the strong dependence of the \ioni{O}{+} abundance on \nel. The 
 simpler explanation is that the \nel([\ion{S}{ii}]) we obtain for the ionization fronts is not indicating the true 
 density of the \ioni{O}{+} zone and, therefore, not appropriate for determining the \ioni{O}{+} abundances. A 
 possible way to check this solution could be to obtain high-spectral resolution spectra to measure and 
 separate the [\ion{O}{ii}] 3726, 3729, 7319 and 7330 \AA\ lines and derive the precise physical conditions 
 of the \ioni{O}{+} zone. Another explanation is that, even in the case that the physical conditions are the 
 same for \ioni{S}{+} and \ioni{O}{+}, \nel([\ion{S}{ii}]) could not be the true one because the reported 
 densities are about --specially in the case of NE-Orion-S-- the critical densities of the [\ion{S}{ii}] 6717, 6731 
 \AA\ lines --2000-3000 \cmc\ and 10000 \cmc, respectively. In fact, considering densities somewhat 
 higher than the true values produce higher \ioni{O}{+}/\ioni{H}{+} ratios. This is because we overestimate 
 the collisional de-excitation affecting the [\ion{O}{ii}] line fluxes. Furthermore, [\ion{O}{ii}] 3726, 3729 \AA\ 
 emission flux can be affected in the same way than [\ion{S}{ii}] lines due to they have similar critical 
 densities --3000 \cmc\ and 5000 \cmc, respectively. Assuming that the main contribution comes from using 
 a wrong density, we have estimated by which amount we have to correct \nel\ in order to obtain the mean 
 nominal value of the O abundance of the Orion Nebula --8.50 dex-- in the zones of our fields with higher 
 \ioni{O}{+}/\ioni{H}{+} ratio. This exercise gives that densities about 1000 and 4000 \cmc\ lower than 
 those determined for the BB and NE-Orion-S fields, respectively, are enough to wash out the problem. These 
 values are rather modest and of the order or even smaller than the uncertainties estimated for 
 \nel([\ion{S}{ii}]) and \nel([\ion{Fe}{iii}]) in each field (\S\ref{dphycon}). Higher signal-to-noise ratio 
 [\ion{Fe}{iii}] spectra and/or good determinations of other indicators with higher critical densities such as 
 \ion{C}{ii}] 2326, 2328 \AA, [\ion{Cl}{iii}] 5517, 5537 \AA\ or even \ion{C}{iii}] 1907, 1909 \AA, as well as 
 the observation of [\ion{O}{ii}] 7319, 7330 \AA\ to derive \ioni{O}{+} abundances would be necessary 
 to disentangle this situation. 
  %%%%%%%%%%%%%
  \begin{table}
   \centering
   \begin{minipage}{60mm}
     \caption{Average distances$^a$ from the peaks of the selected emission lines with respect to 
     that of [\ion{O}{i}]}
     \label{slabs}
    \begin{tabular}{lcc}
     \hline
          & Bright Bar & NE-Orion-S\\     
     \hline
                  [\ion{S}{ii}] &   1\farcs0 $\pm$ 0\farcs4   &  0\farcs4 $\pm$ 0\farcs5 \\
                 
     [\ion{O}{ii}]      &   2\farcs3 $\pm$ 0\farcs9   &  1\farcs4 $\pm$ 1\farcs1\\  
      \ion{H}{i} &   4\farcs5 $\pm$ 1\farcs0   &  3\farcs1 $\pm$ 1\farcs1\\
     \ion{He}{i} &  5\farcs5 $\pm$ 0\farcs9   &  3\farcs3 $\pm$ 0\farcs8\\
 
     [\ion{O}{iii}] &  7\farcs6 $\pm$ 1\farcs1   &  3\farcs9 $\pm$ 0\farcs8\\

     \hline
    \end{tabular}
    \begin{description}
      \item[$^a$] Each value is given as $<x>\pm\sigma$.
    \end{description}    
   \end{minipage}
  \end{table}
  %%%%%%%%%%%%%  
  
 Finally, in Fig.~\ref{f8}, we present the \ioni{O}{2+} and \ioni{C}{2+} abundances derived from RLs for the 
 NE-Orion-S, the only field where these emission lines were satisfactorily detected and measured. From the 
 comparison of Fig.~\ref{f7}e) and Fig.~\ref{f8}a), we can see that the spatial distributions of 
 \ioni{O}{2+}/\ioni{H}{+} determined from CELs and RLs show a rather similar behaviour, placing the 
 maximum abundance values at the spaxels of the northeast corner, where the gas of higher excitation 
 is located. Ionic abundance maps from RLs are nosier than those from CELs due to the faintness of the 
 RLs. As it is common in ionized nebulae, we find that the \ioni{O}{2+} abundances determined from RLs are 
 higher than those from CELs, which is related to the abundance discrepancy problem as we will discuss in 
 \S\ref{adp}
    
%%%%%%%%%%%%%%%%%%%%%%%%%%%%%%%%%%%%%%%%%%%%%%%%%%%%
%%%%%%%%%%%%%%%%%%%%%%%%%%%%%%%%%%%%%%%%%%%%%%%%%%%%
\section{Discussion} \label{discu}
%%%%%%%%%%%%%%%%%%%%%%%%%%%%%%%%%%%%%%%%%%%%%%%%%%%%
 \subsection{Structure of the ionization fronts} \label{struc}
\begin{figure*}
   \centering
   \includegraphics[scale=0.8]{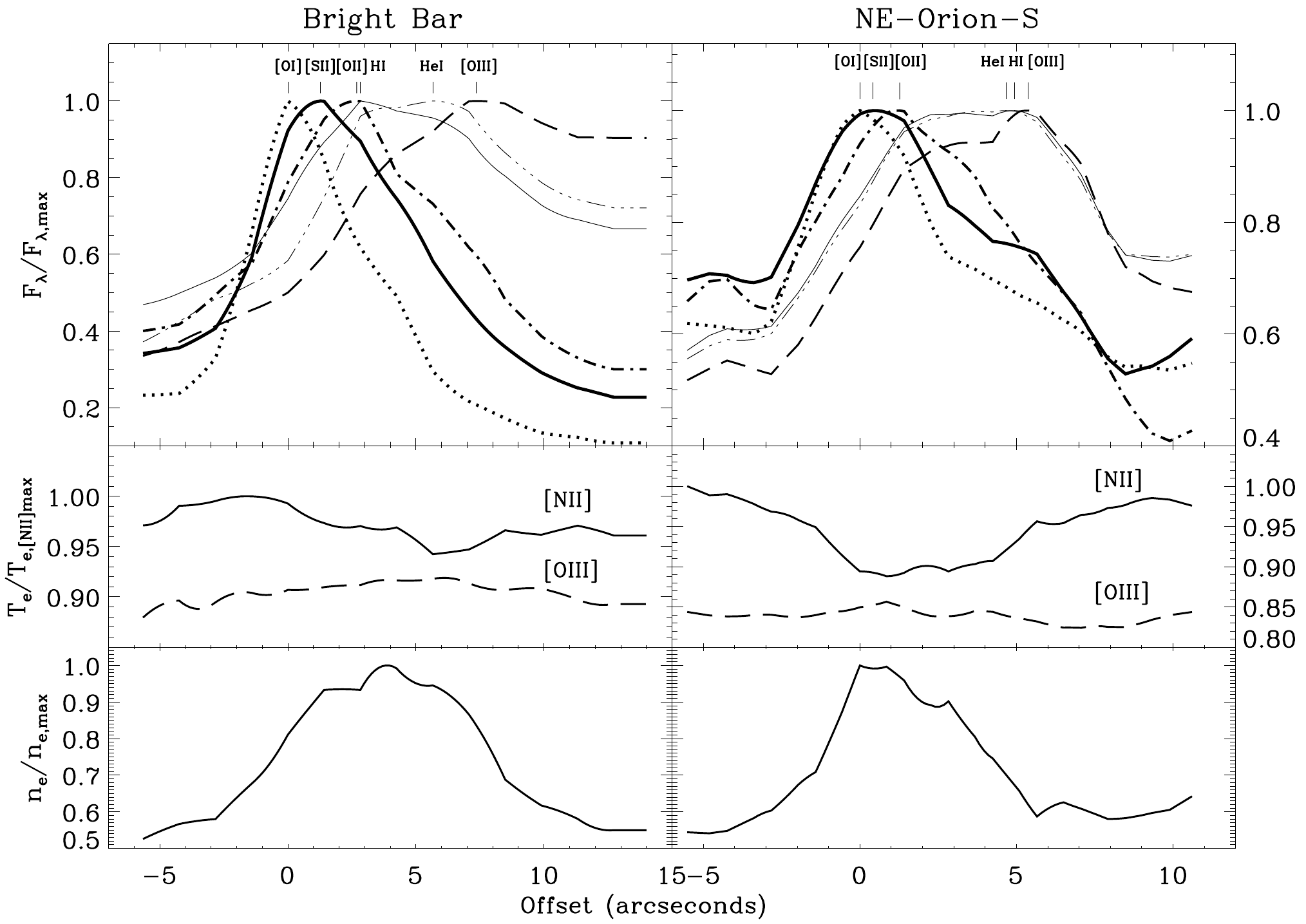} 
   \caption{Spatial profiles of several emission lines traced along the main diagonal from the southeast 
                to the northwest corner of the Bright Bar (top left) and along the third line below (to the southeast) 
                of the main diagonal from the southwest to the northeast corner of the NE-Orion-S (top right). 
                The zero point is located at the maximum of [\ion{O}{i}] line emission and the offsets are given in 
                arcseconds on the plane of the sky. Positive values in abscissae indicate decreasing distance 
                with respect to $\theta^1$Ori C. The plotted lines are: [\ion{O}{i}] (dotted line), [\ion{S}{i}] 
                (black solid line), [\ion{O}{ii}] (dashed-dotted line), \ion{H}{i} (thinner solid line), \ion{He}{i} 
                (three dotted-dashed line) and [\ion{O}{iii}] (dashed line). The emission line fluxes are normalized 
                to their maximum emission. In the middle and bottom panels, the spatial profiles of 
                \te([\ion{N}{ii}]) (solid line) and \te([\ion{O}{iii}]) (dashed line), as well as \nel([\ion{S}{ii}]) 
                are presented. \te([\ion{N}{ii}]) and \nel([\ion{S}{ii}]) are normalized to their respective maxima: 
                9480 K and 4760 \cmc\ for the Bright Bar and 10390 K and 14200 
                \cmc\ for the NE-Orion-S.}
   \label{ioni}
  \end{figure*}

  In this section we study the ionization fronts of the two observed fields. With this aim, we analyse the spatial 
  profiles of emission lines emitted from ions with different ionization potentials (\ioni{O}{0}, \ioni{S}{+}, 
  \ioni{O}{+}, \ioni{H}{+}, \ioni{He}{+} and \ioni{O}{2+}). We have extracted the spatial profiles from cuts 
  roughly parallel to the radial direction to $\theta^1$Ori C, connecting the southeast with the northwest 
  corner in the BB field and the southwest with the northeast corner in the NE-Orion-S field. The use of all 
  possible profiles along these directions allowed us to statistically determine the relative distance 
  between the maximum of the spatial distribution of the [\ion{O}{i}] emission and the other emission 
  lines across the fields. To estimate these distances, we have used the rebinned maps without applying 
  any Gaussian smoothing filter as we did for the H$\alpha$ images presented in Fig.~\ref{f1}. The 
  rebinned maps do not change the relative positions of the maxima and allow us to explore 
  the changes in the profiles interpolating at subarcsecond scales. Furthermore, these maps allow to use 
  a total of 29 spatial profiles to calculate statistically significant averages. In Table~\ref{slabs} we 
  present the average values of these distances and their standard deviations. In Fig.~\ref{ioni} we 
  present spatial profiles of the emission lines only for one representative cut: along the main diagonal in 
  the case of the BB; and along the third diagonal line towards the southeast of the main diagonal in the 
  case of the NE-Orion-S. In Fig.~\ref{ioni} we also show the temperature and density distributions along the 
  same cuts. For all profiles, we have located the zero point at the position of the maximum of 
  [\ion{O}{i}] emission, which is the farthest one from $\theta^1$Ori C, as we saw in Fig.~\ref{f4} and see now 
  in Fig.~\ref{ioni}.
  %%%%%%%%%%%%%%% 
 %%%%%%%%%%%%%%%
 
  Theory about the ionization structure of nebulae can be found in \cite{osterbrockferland06}. In our 
  discussion below, we state that the stratification should be driven by the ionization energy of the 
  atom. This is approximately true, but it should also be mentioned that the local ionizing radiation field in a 
  nebula, which control the distribution of the ionic states of heavy elements, is determined by the 
  opacity of the two most abundant elements, hydrogen and helium. In terms of the emission lines 
  fluxes presented in Fig.~\ref{ioni}, this implies that both [\ion{O}{i}] and [\ion{S}{ii}] arise near the zone 
  where hydrogen becomes optically thick and the presence of these ions is combined with thermalized 
  electrons coming from hydrogen photoionizations. [\ion{O}{ii}] as well as [\ion{N}{ii}] emissions arise from 
  the zone where hydrogen is ionized and helium is still neutral. Finally, [\ion{O}{iii}] emission comes from 
  the nearest zone to $\theta^1$Ori C, where hydrogen remains ionized and helium is once ionized. In the 
  spatial profiles presented in Fig.~\ref{ioni}, the small differences with respect to that expected stratification 
  are probably a combined effect of the geometry of each ionization front and the density dependence of the 
  observed flux.     
  
  The ionization structure of the BB has been well studied in the literature \citep[$e.g.$][]{hester91,
  tielensetal93, walmsleyetal00,pellegrinietal09,vanderwieletal09}. This bar is basically edge-on and its 
  plane has an inclination of 7$^\circ$ with respect to the line of sight \citep{pellegrinietal09}, allowing to 
  resolve its ionization structure as we see in Fig.~\ref{ioni}. The MIF is a thin layer located between the 
  maximum emission peaks of [\ion{O}{i}] and [\ion{S}{ii}]. \cite{pellegrinietal09} indicate that the maxima of 
  H$_2$ and $^{12}$CO line emissions are detected at about 12$\arcsec$ and 20$\arcsec$ behind the MIF 
  position, respectively. As we see in Table~\ref{slabs}, we found that the peak of the [\ion{S}{ii}] emission is 
  located about 1$\arcsec$ ahead [\ion{O}{i}], which corresponds to a physical separation of 
  2.1$\times10^{-3}$ pc assuming a distance to the Orion Nebula of 436$\pm$20 pc \citep{odellhenney08}. 
  The maximum of [\ion{O}{ii}] emission, which also coincides with the maxima of [\ion{N}{ii}] (14.5 eV) and 
  [\ion{Fe}{iii}] (16.2 eV), is found about 2\farcs3 --4.9$\times10^{-3}$ pc-- with a standard deviation of 
  0\farcs9, implying that this zone is spatially less defined than that of [\ion{S}{ii}] emission. The spatial 
  profiles of the \ion{H}{i} and \ion{He}{i} emission lines are more extended. The mean distance of the peak of 
  \ion{H}{i} emission is 4\farcs5 (9.5$\times10^{-3}$ pc) with a standard deviation of 1$\arcsec$. In the case 
  of \ion{He}{i} emission, its maximum is located between the \ion{H}{i} and [\ion{O}{iii}] zones --as it is 
  expected considering the ionization potential of this ion (24.6 eV)-- at 5\farcs5 (11.6$\times10^{-3}$ pc) 
  with respect to the peak of [\ion{O}{i}] emission. The \ion{He}{i} emission distribution represented in 
  Fig.~\ref{ioni} corresponds to the line at 6678 \AA, which permits to avoid problems related to 
  self-absorption due to it comes from a singlet state. Finally, the [\ion{O}{iii}] zone reaches its maximum 
  value at 7\farcs6 (16.1$\times10^{-3}$ pc).   
   
  In contrast to the BB, the ionization structure of NE-Orion-S has not been described in the literature. Taking 
  into account the distances estimated in Table~\ref{slabs}, the distribution of the spatial profiles shown 
  in Fig.~\ref{ioni} and comparing with the results for the BB, it is clear that we can not resolve the 
  ionization structure of the NE-Orion-S as well as of the BB because this front is much tilted with respect to 
  the line of sight. In their figure 5b, \cite{wenodell95} present the geometry of their Orion Nebula model along 
  a cut with a position angle of 226$^\circ$, which passes close 
  to the southeast corner of our NE-Orion-S field. At a distance of 30$\arcsec$ from the center of our field, the 
  model by \cite{wenodell95} indicates that the MIF has an inclination of about 60$^\circ$ with respect to 
  the line of sight and after that the MIF remains rather parallel to the plane of the sky. The properties of 
  the nebular structure to whom NE-Orion-S belongs, the Orion-S cloud, have been recently reviewed by 
  \cite{odelletal09} showing that it should be located in the foreground with an unknown inclination as well 
  as an uncertain placement in the three-dimensional space, but at a similar distance with respect to 
  $\theta^1$Ori C as that of the MIF in the background (0.182 pc). The inclination angle of the NE-Orion-S 
  have to be such to produce the observed distances between line emission maxima given in 
  Table~\ref{slabs}, where, on the one hand, the emission of [\ion{S}{ii}] and [\ion{O}{ii}] and, on the 
  other hand, the emissions of \ion{H}{i}, \ion{He}{i} and [\ion{O}{iii}] are not well separated at our spatial 
  resolution. We can compare the data of Table~\ref{slabs} for the BB and NE-Orion-S to obtain a first order 
  estimation of the NE-Orion-S inclination. Assuming a proportional relation between the bar inclination with 
  respect to the plane of the sky and the average distances with respect to the [\ion{O}{i}] maximum emission 
  of Table~\ref{slabs}, we have estimated that the plane of the NE-Orion-S should be tilted about 
  48$^\circ\pm$13$^\circ$ with respect to the line of sight. This angle was calculated as a weighted average of 
  the individual estimations for each emitting zone. In addition, it is important to note that the NE-Orion-S has 
  a complex structure of shadowed regions and other features as we see in Fig.~\ref{f1} as well as in the new 
  combined image of the Huygens region obtained by J.R. Walsh using a $drizzle$ technique to produce 
  a mosaic with a resolution of 0\farcs1 over the whole field\footnote{J.R. Walsh webpage: 
  http://www.stecf.org/~jwalsh/}. Furthermore, the proximity and unknown exact position with respect to 
  the Trapezium cluster is also playing a role on the geometry of what we observe as we mention in 
  \S\ref{elf}. Therefore, it can be difficult to interpret this ionization front as a simple slab with a given 
  inclination. In this sense, our analysis of the individual spatial profiles extracted parallel to the main 
  diagonal of the NE-Orion-S field can give further information about the complex structure of the 
  NE-Orion-S. We have noted that the emission of [\ion{S}{ii}] and [\ion{O}{ii}] are much separated in the 
  profiles extracted below --to the southeast of-- the main diagonal than above it. In fact, we have 
  estimated that the average distances between the emission peaks of [\ion{S}{ii}] and [\ion{O}{ii}] and that 
  of [\ion{O}{i}] are of about 0\farcs7$\pm$0\farcs4 and 1\farcs9$\pm$0\farcs9 below the main diagonal, 
  and 0\farcs1$\pm$0\farcs3 and 0\farcs6$\pm$0\farcs3 above it. This fact suggests that the inclination of 
  the NE-Orion-S is not constant along the observed field. Below the main diagonal, the plane of the 
  NE-Orion-S seems to be less inclined with respect to the line of sight --more edge-on. But, above the 
  main diagonal, the inclination angle is larger and the peaks of the line emission of the different ionic species 
  can not be resolved. The variation of the inclination of the NE-Orion-S along the field may also explain 
  the change of the width of the area of high [\ion{N}{ii}]/H$\alpha$ ratio --a good indicator of the inclination of 
  an ionization front, as it is commented in \S\ref{elf}-- across the field (see Fig.~\ref{f5}), which seems to be 
  wider below the main diagonal.  
  
  In Fig.~\ref{ioni} we have also plotted the spatial profiles of physical conditions of the BB and NE-Orion-S 
  along the same diagonals as in the case of upper panels. Here, some of the results commented in 
  \S\ref{phycon} are more clearly seen: \te([\ion{N}{ii}]) is always higher than \te([\ion{O}{iii}]), but they tend 
  to be more similar inside te bars. For the NE-Orion-S we can note a correlation between the \nel\ and 
  \te([\ion{N}{ii}]) profiles, where lower temperatures are reached where the densities are higher. At the BB, 
  this effect is not as clear as in the case of NE-Orion-S, this can be due to the lower densities of the BB.      
%%%%%%%%%%%%%%%%%%%%%%%%%%%%%%%%%%%%%%%%%%%%%%%%%%%%
 \subsection{Abundance discrepancy problem} \label{adp}  
 The Orion Nebula is an unique object to analyse and explore particular problems in common with other 
 \hii\ regions. This is the case of the well known abundance discrepancy (AD) problem, $i.e.$ the 
 disagreement between the abundances of the same heavy-element ion derived from CELs and RLs. The 
 AD problem is quantified making use of the AD factor, ADF, which can be defined for a given ion as  
 \begin{equation}
    {\rm ADF(X^{i+})} = log\Big({\rm \frac{X^{i+}}{H^+}}\Big)_{RL} - 
                                  log\Big({\rm \frac{X^{i+}}{H^+}}\Big)_{CEL}.
  \end{equation}   

 From intermediate and high resolution spectroscopy, several works have found that the \adfo\ is always 
 between 0.1 and 0.3 dex in Galactic and extragalactic \hii\ regions \citep[see $e.g.$][]{estebanetal02, 
 tsamisetal03, estebanetal04, garciarojasesteban07, lopezsanchezetal07,mesadelgadoetal08, 
 estebanetal09, mesadelgadoesteban10}. \cite{garciarojasesteban07} argued that the AD problem in \hii\ 
 regions seems to be consistent with the predictions of the temperature fluctuation paradigm proposed by 
 \cite{peimbert67} and parametrized by the mean square of the spatial variations of temperature, the 
 so-called \tf\ parameter. Under this hypothesis, the AD would be produced by the very different 
 temperature dependence of the emissivities of both kinds of lines and the abundances from RLs would be 
 the representative ones of the ionized gas. However, the existence and origin of such temperature 
 fluctuations are still controversial and the AD problem remains open. Other authors as 
 \cite{tsamispequignot05} and \cite{stasinskaetal07} have proposed 
   %%%%%%%%%%%%%%% 
   \begin{figure}
   \centering
   \includegraphics[scale=1.3]{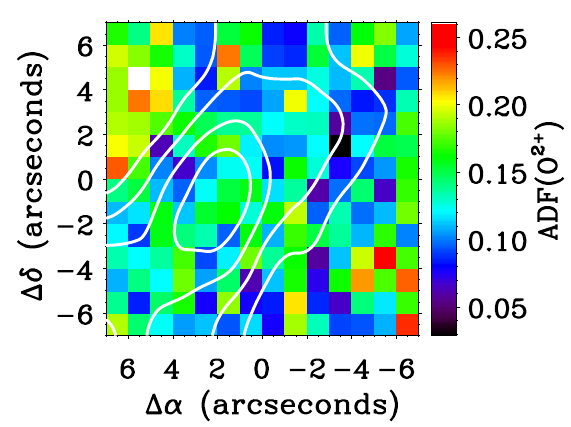} 
   \caption{Spatial distribution of \adfo\ for the NE-Orion-S with H$\alpha$ contours overplotted.}
   \label{f10}
  \end{figure}
 %%%%%%%%%%%%%%%  
 a different hypothesis to explain the AD problem in \hii\ regions. Based on the model for heavy-element 
 mixing of \cite{tenoriotagle96}, \cite{stasinskaetal07} propose that the presence of cold metal-rich droplets of 
 supernova ejecta still not mixed with the ambient gas of the \hii\ regions would produce most of the RL 
 emission, while CEL emission would be produced by the ambient gas, which would have  the expected 
 composition of the local interstellar medium. In such case, the abundances from CELs would be the truly 
 representative ones of the nebula. Recently, the studies performed in the Orion Nebula by 
 \cite{mesadelgadoetal09b}, \cite{tsamisetal11} and \cite{simondiazstasinska11} have provided new 
 observational clues to address this puzzling issue. \cite{mesadelgadoetal09b} obtained echelle 
 spectroscopy of the HH~202-S separating the kinematic component of the background emission from that  
 of the gas flow finding clearly higher \adfo\ values at the high-velocity component. This result suggests 
 a possible relation between the \adfo\ and the gas flow velocity. \cite{tsamisetal11} carried out a 
 high-spectral resolution analysis of the protoplanetary disc LV2 with FLAMES. Their study of LV2 
 emission with the nebular background subtracted shows \adfo\ values which are basically zero. This 
 surprising result seems to be related to the high electron densities expected in these kind of objects, which 
 affect the ionic abundances derived from CELs. Therefore, it is necessary to explore the behaviour of the 
 \adfo\ in other morphological structures of the Orion Nebula --as stratified ionization fronts-- in order to 
 better understand and constrain the AD problem. Finally, we consider important to comment the recent 
 remarkable result by \cite{simondiazstasinska11}, who find that the nebular oxygen total abundances 
 derived from RLs agrees better with the stellar abundances of B-type stars of the Orion Association than 
 those determined from CELs. This result is an indication that nebular abundances obtained from RLs may 
 be more reliable than those derived from the standard method based on CELs and therefore stressing the 
 importance of investigating the AD problem.  
  
 As we pointed out in \S\ref{cheab} we derived \ioni{O}{2+} abundances from RLs in the NE-Orion-S field. 
 Subtracting these abundances from those obtained from CELs (Fig.~\ref{f7}) we find the spatial distribution 
 of the \adfo, which is presented in Fig.~\ref{f10}. As we can see, the \adfo\ seems to be slightly higher at 
 the northeast corner of the field, but the map does not show any remarkable structure inside the ionization 
 front. The range of the variations goes from 0.05 to 0.25 dex with a typical error of about 0.12 dex for each 
 spaxel. However, we have checked that the extreme values of the \adfo\ correspond typically with 
 spaxels with low signal-to-noise ratio in the \ion{O}{ii} RLs of their spectra. Therefore, not particular 
 variations of the \adfo\ are related to the presence of stratified ionization fronts, at least with the 
 properties and physical conditions of the NE-Orion-S and the spatial resolution of our observations. The 
 mean value of the \adfo\ in the field is about 0.14 dex with a standard deviation of about 0.05 dex. Therefore, 
 there is a real difference between the \ioni{O}{2+} abundances from RLs and CELs, which is in the range 
 between 21\% and 44\%. This is precisely the value found in our previous studies of the Orion Nebula 
 \cite[$e.g.$][]{estebanetal04,mesadelgadoetal08}. 
 
%%%%%%%%%%%%%%%%%%%%%%%%%%%%%%%%%%%%%%%%%%%%%%%%%%%%
%%%%%%%%%%%%%%%%%%%%%%%%%%%%%%%%%%%%%%%%%%%%%%%%%%%%
\section{Conclusions} \label{conclu}
 In this paper we analyse integral field optical spectroscopy of two two remarkable structures related to 
 ionization fronts of the Orion Nebula: the Bright Bar (BB) and the northeast edge of the Orion South 
 cloud (NE-Orion-S). In both fields, we have mapped emission line fluxes and ratios, physical conditions 
 and chemical abundances at spatial scales of 1$\arcsec$. 
 
 The maps clearly show the ionization stratification of both ionization fronts, which present rather different 
 inclination angles as well as different distances with respect to $\theta^1$Ori C, the main exciting source of 
 the nebula. The ionization structure of the BB is well resolved because the ionization front is almost 
 edge-on --inclination angle of 7$^\circ$ with respect to the line of sight--, while the NE-Orion-S 
 is a portion of the Orion-S cloud suspended within the main body of the nebula, which is more tilted and 
 presents a more complex structure. We have analysed spatial profiles of selected emission lines of ions 
 with different ionization potential and estimated the average distances of the peaks of the emission of 
 these lines with respect to the maximum of [\ion{O}{i}]. Comparing those distances in both fronts, we 
 have estimated that the plane of the NE-Orion-S field could be tilted about 48$^\circ\pm$13$^\circ$ 
 with respect to the line of sight. However, the distances between the peaks of the lines with respect to 
 [\ion{O}{i}] change in the different spatial profiles extracted for the NE-Orion-S, indicating that the inclination 
 angle of this ionization front should vary across the observed field.  
  
 The maps of electron density, \nel(\ion{S}{ii}]), show that NE-Orion-S is much denser than the BB. The 
 NE-Orion-S has a mean density of about 10900 \cmc, while the BB has 4370 \cmc. In both cases, the 
 [\ion{S}{ii}] lines may be somewhat affected by collisional de-excitation because the nominal values of 
 \nel\ we find are of the order of their critical densities. In any case, this de-excitation should not be severe 
 due to the rather similar \nel\ we determine from [\ion{Fe}{iii}] lines. The structure of the temperature 
 maps are rather featureless and \te([\ion{N}{ii}]) is usually higher than \te([\ion{O}{iii}]), specially outside 
 the bars. The maps of total O abundance show mean values consistent with previous determinations of the 
 literature for the Orion Nebula, but they show a pattern very similar to the maps of \ioni{O}{+}/\ioni{H}{+} 
 ratio and \nel([\ion{S}{ii}]). The spaxels with higher \nel\ values also show higher \ioni{O}{+} and O 
 abundances. We find that considering values of \nel\ slightly lower than those determined from the 
 [\ion{S}{ii}] line ratio --but of the order of the quoted uncertainties-- the problem can be satisfactorily solved. 

 Finally, the higher signal-to-noise ratio of the spectra in the NE-Orion-S field allows us to detect, measure 
 and map the faint RLs of \ion{O}{ii} and \ion{C}{ii}. We explore the abundance discrepancy problem in this 
 field comparing the \ioni{O}{2+} abundance maps derived from CELs and RLs, finding a featureless spatial 
 distribution for the abundance discrepancy factor of \ioni{O}{2+}, \adfo, which shows values ranging 
 between 0.05 and 0.25 dex with a typical error of about 0.12 dex. The mean \adfo\ is 0.14$\pm$0.05 dex, 
 in excellent agreement with other previous determinations in other areas of the Orion Nebula --except at 
 Herbig-Haro objects \citep{mesadelgadoetal08,mesadelgadoetal09a} and the LV2 protoplanetary 
 disc \citep{tsamisetal11}. This result implies that the \ioni{O}{2+} abundances from RLs are between 
 21\% and 44\% larger than those from CELs and that there is not any particular variation of the \adfo\ 
 related to the presence of stratified ionization fronts, at least for the properties and physical conditions of 
 the NE-Orion-S and the spatial resolution of our observations
    
%%%%%%%%%%%%%%%%%%%%%%%%%%%%%%%%%%%%%%%%%%%%%%%%%%%%
%%%%%%%%%%%%%%%%%%%%%%%%%%%%%%%%%%%%%%%%%%%%%%%%%%%%
\section*{Acknowledgments} 
We are very grateful to the referee of this paper, C.R. O'Dell, for his useful comments and suggestions. 
We also thank W. Henney, C. Morriset and Y. Tsamis their helpful comments and correspondence. This work has 
been funded by the Ministerio de Educaci\'on y Ciencia (MEC) under project AYA2007-63030.

%%%%%%%%%%%%%%%%%%%%%%%%%%%%%%%%%%%%%%%%%%%%%%%%%%%%
%%%%%%%%%%%%%%%%%%%%%%%%%%%%%%%%%%%%%%%%%%%%%%%%%%%%
%\bibliographystyle{mn2e}
%\bibliography{mnrasmnemonic,biblio}

%%%%%%%%%%%%%%%%%%%%%%%%%%%%%%%%%%%%%%%%%%%%%%%%%%%%
%%%%%%%%%%%%%%%%%%%%%%%%%%%%%%%%%%%%%%%%%%%%%%%%%%%%

\label{lastpage}

\end{document}